\documentclass[preprint2]{aastex}

\shorttitle{Merging of globular clusters}
\shortauthors{Miocchi et al.}

\def\beq{\begin{equation}}
\def\eeq{\end{equation}}

\def\mean#1{<\!\!#1\!\!>}
\def\a{(a) }
\def\b{(b) }
\def\c{(c) }
\def\d{(d) }

\begin{document}

\title{Merging of globular clusters within inner
galactic regions.\\
I. Do they survive the tidal interaction?
}

\author{P. Miocchi}
\email{miocchi@uniroma1.it}
\author{R. Capuzzo Dolcetta}
\email{dolcetta@uniroma1.it}
\author{P. Di Matteo}
\email{p.dimatteo@uniroma1.it}
\and
\author{A. Vicari}
\email{alessandrovicari@uniroma1.it}
\affil{Dipartimento di Fisica, Universit\'a di Roma ``La Sapienza",\\
P.le Aldo Moro, 2, I00185 -- Rome, Italy.}

\begin{abstract}
The main topic of this paper is the investigation
of the modes of interaction of globular clusters (GCs) moving
in the inner part of a galaxy.
This is tackled by means of high-resolution $N$-body simulations,
whose first results are presented in this article.
Our simulations dealt with 
primordial very massive  
(order of $10^7$ M$_\odot$) GCs
that were able to decay, because of dynamical friction,
into the inner regions of triaxial galaxies on
a time much shorter than their internal relaxation time.
To check the disruptive roles of both tidal forces and
GC-GC collisions, their effects were 
maximised by considering clusters on quasi-radial orbits and
choosing the initial conditions so as to give head-on
collisions at each passage through the center.

The available CPU resources allowed us to simulate clusters
with different structural parameters and to follow them on quasi-radial
orbits during 8 passages across the center. The main findings are:
i) clusters with an initial high enough King concentration
parameter ($c\geq 1.2$), preserve up to 50\% of their initial mass;
ii) the inner density distribution of the survived clusters keep a King model profile;
iii) GC--GC collisions have a negligible
effect with respect to that caused by the passage through the galactic center;
iv) the orbital energy dissipation due to the tidal interaction is of the same order
of that caused by dynamical friction; v)
complex sub-structures like ``ripples'' and
``clumps'' formed, as observed around real clusters.
These findings support the validity of the hypothesis of merging
of GCs in the galactic central region, with modes that deserve further
careful investigations.
\end{abstract}


\section{Introduction}
The study of the detailed structure and evolution of globular clusters (GCs)
in galaxies is a modern astrophysical concern, which is
particularly suitable for HST \citep{k99} and for large ground-based
telescopes \citep{g01}.
Some interesting general considerations on the structural properties of
GCSs in galaxies are found in \citet{az97}, where
also a summary of simplified models and results by other authors on the
topic of formation and destruction of globular clusters in galaxies is present.
While the amount of data for GCSs in galaxies is
now large and rapidly increasing \citep[see, e.g.][]{davidge, dirsch, forbes, fort, gomez,
k04,olsen,zepf},
from a theoretical point of view the study of the evolution of GCSs
in galaxies has been tackled in different ways but it still lacks of
definitive conclusions.
We just recall here the studies by \citet{aho88}, \citet{go97} and
\citet{v01}, which aim at achieving
a deeper understanding of the effects of the interaction of clusters with the
galactic field, and the interesting results obtained by \citet{gnedin99} on the survival
of clusters in dependence on their initial conditions.

It is known that, in elliptical galaxies, the most important
processes causing the evolution of GCs are i) the tidal interaction with
the global field \citep{muralia}, that has also ii) the effect of accelerating the
relaxation evaporation \citep{muralib}; iii) the impulsive interaction with a
compact object in the galactic center (on much smaller
length-scales), see \citet{obs89}, \citet{bib3}, \citet{ct97}, \citet{ct99};
iv) the dynamical friction (df) due to the 
bulge-halo matter as firstly shown by
\citet{tos75} and then generalised to the triaxial galaxies case by
\citet{pesce}, \citet{bib3}, \citet{cv04}, who ascertained that df plays
a much more important role
than what previously believed; see also \citet{colpi99}, \citet{vandenbosch99},
\citet{penarrubia02}.

Actually, \citet{pesce} (hereafter PCV), \citet{bib3} and \citet{cv04} showed
quantitatively (by mean of numerical integration of a large set of orbits
of GCs in self-consistent galactic models) that massive
enough clusters have short orbital decaying times.
More specifically, Eqs. (A1) and (A2) in \citet{bib3} indicate the (almost)
total loss of the orbital energy in less than 5 Gyr for $10^6$
M$_\odot$ GCs  on low angular momentum orbits having an initial orbital binding
energy per unit mass $E=0.6$,  which is  (in units of the galactic total potential depth)
the energy corresponding to the background stellar velocity dispersion in
the assumed Schwarzschild's triaxial model \citep{schw}; this decay time,
inversely scaling with GC mass,
is less than $500$ Myr for globulars more massive than
$10^7$ M$_\odot$. 
Incidentally, the apocentric distance of a quasi-radial orbit (thin box)  of $E=0.6$ is
$6.3r_b$ while the radius of the quasi-circular orbit of same energy is $3.8r_b$,
where $r_b$ is the core radius of the Schwarzschild's (1979) model spherical
component; the assumption (as in Schwarzschild 1979) $r_b=200$ pc
gives $1.3$ kpc and $760$ pc for the apocenter and circular radius, respectively.
The more recent paper \citet{cv04} represents a generalization and widening
of the previous PCV and \citet{bib3} works and refers to a set of model
for triaxial galaxies with a density core and various axial ratios.
For the sake of comparison with the above cited results based
on the Schwarzschild's model,
a semimajor axis of 300 pc and the same axial ratios 2:1.25:1 of that
model are used in the df decay time fitting formulas given in
\citet{cv04} to get, for the same $10^6$ M$_\odot$ GC of the same initial
orbital energy and angular momentum,
a decay time  $10^9$ yr, i.e. a factor $5$ shorter than the one
evaluated as described above. Being the df time easily evaluated for any
GC mass, due to its inverse proportionality to the mass, we get a decay time 
ranging from $10^8$ yr to $20$ Gyr for an interval of GC masses extending from
$5\times 10^4$ M$_\odot$ to $10^7$ M$_\odot$, implying a significant evolution of an
initial GC population by df, as actually shown  by \citet{cv04}.

One application of the present work is to the study of the validity
 of an 
intriguing scenario elsewhere presented \citep{bib3} and that naturally 
emerges 
from all these considerations: 
the high efficiency of the df implies that many massive 
($M \gtrsim 5\times 10^6$
M$_\odot$) globular clusters decay rapidly to the inner galactic region where they 
interact closely with each other and with the galactic nucleus, eventually forming, 
through a merging process, a dense and
massive super-cluster.
This scenario, that requires an accurate modeling, 
has important implications both on the accretion of massive
objects in the galactic center and on their radiative emission as AGN 
\citep[see the discussion in ][]{bib3,cv04}.

One crucial point for this scheme to work is the existence
of few primordial massive GCs. At this regard, we remind that the real 
problem was, once, to understand why very massive GCs were 
not observed, indeed. This because the typical Jeans 
mass in a primordial, virialized, gaseous galactic halo ($T\simeq 10^5 \div 10^6$ K)
is $>10^7$ M$_\odot$, which is too large respect to the observed (Milky Way)
GC mass values. This led to the theoretical attempt to 
understand why such massive GCs were not observed, on the basis of various schemes and scenarios 
of GC formation (\citet{fallrees},\citet{vietripesce})
mainly invoking the efficience of cooling mechanisms reducing the temperature 
to low enough values. The new, relevant, point is that there are now quite a
few recent papers showing, indeed, the existence of \emph{\/young} massive 
clusters in Antennae \citep{fritze}, in the Magellanic Clouds, M33, Fornax dSph \citep{degr}.
Young massive clusters have been found, also, in M31 \citep{fusip};
in this galaxy, \citet{huxor} discovered bright clusters with
anomalously large half-mass radii ($\approx 30$ pc).

In particular, \citet{fritze} showed that the mass function
of young clusters in the Antennae extends up to few times $10^7$ M$_\odot$.
The existence of very massive GCs does not seem to be a peculiarity of young
systems, only. Actually,  \citet{harpud} give the evidence of the presence of
$10^7$ M$_\odot$  in the giant elliptical M87 as well as in Virgo
ellipticals, being the cumulative (sum over 3 ellipticals in the cluster) mass
function of GCs extended up to high masses.
The observed  presence of very massive GCs fits with the recent
theoretical-numerical findings  by \citet{kravt} who deduce a dependence
of the most massive GC mass ($M_{max}$) on the parent galaxy mass ($M_g$),
$M_{max}\propto M_g^{1.29}$, such that $M_{max} \simeq 10^7$ M$_\odot$ for
$M_g \simeq 2.6 \times 10^{11}$ M$_\odot$.
Moreover, they also find that very massive GCs
contribute to more than $50\%$ to the total cluster mass, in fine agreement
with the observational data of \citet{harris06} that indicate how up a full
$40\%$  of the total mass that is now in the GCSs of brightest cluster
galaxies is contributed by massive (present day mass
$> 1.5 \times 10^6$ M$_\odot$).
It may be also worth remembered the specific paper by \citet{baumgetal} devoted
to the modeling of G1, the massive GC in M31, obtaining for it a mass
$8\times 10^6$ M$_\odot$.

Consequently, one can just make a hypothesis on the initial abundance
of massive clusters and see whether it gives results consistent with 
available observations, checking, of course the dependence of 
results on the assumption.
This was done in \citet{cv04}, that shows (see their Fig.11) 
how few tens of massive ($M>10^7$ M$_\odot$) clusters suffice
to give an accretion rate onto a central galactic 
black hole high enough (few M$_\odot$ yr$^{-1}$) 
to sustain an AGN activity.
Given all this,
our main concern is to check whether or not the tidal distortion suffered
by the GCs due to the halo and the bulge is destructive on the time-scale
needed by df to dissipate the cluster orbital energy. The detailed
numerical studies performed by \citet{laguna95,nord99} give encouraging support to
this claiming.
In this paper, we show the results of detailed $N$-body simulations that take into
account df and two other concurrent effects: i) the tidal interaction of a GC with the overall galactic
field and ii) the collision with another passing-by GC.
We consider quasi-radial orbits for GCs,
such to maximise the tidal effects.

As a `by-product' of this work, tidal tail formation and morphology is analysed and a
particular attention is focused on the presence of overdensities (clumps)
in the tails.
Similar sub-structures have been detected in the two tidal tails
of the globular cluster Palomar 5 \citep{oden01,oden03},
while the presence of tails surrounding many other GCs is strongly suggested by various
observations \citep{ls97, testa00, lmc00, sieg00, lee03}.
NGC 6254 and Palomar 12 seem to show similar overdensities in their tails,
too \citep{lmc00}.
Various authors attribute the formation of clumps in the tails to strong
gravitational shocks suffered by the cluster \citep{combes99} and, in particular, to the
tidal effect due to compact galactic sub-structures \citep{dehn04}, like molecular clouds
or spiral arms.
Nevertheless, the mechanisms of their origin and formation is still unclear and
deserve an adequate interpretation.
More recently, Capuzzo Dolcetta, Di Matteo \& Miocchi (2005)
(hereafter CDM) showed that clumpy tails
emerge also in GCs moving in a regular `smooth' galactic environment on both quasi-circular and
more eccentric orbits, suggesting that such clumpy sub-structures are related to the
local decrease of the stars velocity along the tails (see also
Di Matteo, Capuzzo Dolcetta \& Miocchi, 2005).

This paper is organized as follows: in Sect.\ref{model} the numerical
modeling is described, the results are shown and discussed in Sect.\ref{results}
and conclusions drawn in Sect.\ref{concl}.

\section{The model}
\label{model}
We consider GCs as $N$-body systems moving within a triaxial
galaxy represented by an analytical potential (see next Section),
including the deceleration due to dynamical friction caused by 
the field stars.
We studied the evolution of a pair of GCs
whose center-of-mass (CM) initial conditions were assigned such to put
them along very elongated orbits.
They start moving on opposite sides respect to the galactic center
and oscillate quasi-radially around it, crossing and colliding each other
in the galactic center neighbourhood (see, e.g., the upper panel
in Fig.~\ref{energiaB}).
Both the choice of very elongated orbits and the presence of two clusters
were motivated by the wish to study the combined effects of the tidal interaction and
of the collisions between GCs. Indeed, when such systems undergo
the final stages of orbital decay, GC-GC merging and close interactions are
expected to occur.

We dealt with four different types of clusters, named \a, \b, \c and \d
at increasing order of concentration (see Table \ref{tab1} and Sect.\ref{clumod}).
Their dynamics was modelled in simulation A, that involved
the pair of clusters \a and \b, and in simulation B that regarded the
more compact clusters \c and \d.
We also considered a further case (simulation C) where the dynamics of cluster \a,
with the same initial conditions as in simulation A, was followed \emph{\/without} the
presence of the other cluster.
This latter simulation lasted less than the others due to CPU-time limitations.
Note that the total CPU-time spent by all the simulations presented
here is about 30,000 hours divided among the 64 processors used on an IBM SP4
system (granted by the INAF-CINECA agreement).

In this paper, unless otherwise specified, lengths, masses and time are
measured, respectively, in unit of the galactic core radius $r_b$, of the
galactic core mass $M_b$ and of the galactic core crossing time $t_b=(r_b^3/GM_b)^{1/2}$.
Note that the numerical
values in physical units of the mentioned parameters are irrelevant for the
results of the simulations, that can be always scaled as long as
the ratios $t/t_b$, $r/r_b$, $M/M_b$ (with $M$ the total cluster mass)
are kept unchanged.
This is strictly true only if one can neglect, like in our case, the effects
due to 2-body stellar collisions (as discussed in Sect.\ref{clumod}), otherwise
the dynamics would depend also on the mass of the \emph{\/single} star in the clusters.

The reference frame was fixed with the origin at the galactic center and
the $x$ and $z$ axes, respectively, along the maximum and minimum axis of
the triaxial ellipsoid.
The cluster CM was initially located on the $x$-axis: cluster \a
and \c at $x_0=-4.15$, and cluster \b and \d at $x_0=3.95$.
In all the cases the initial velocity components were $(0,0.05,0.025)$.
Because of the CPU time limitations, we were not able to consider a wide set
of initial conditions, thus we decided to choose initial velocity and position
in such a way as to set \emph{\/upper} limits to the disruptive
effects exerted on massive clusters by the tidal interaction with the
galactic field and by the close interaction with other GCs.
For this reason, we chose almost
radial orbits so that the GCs cross twice per period the galactic
center, where the tidal interaction is strongest, and
they undergo head-on collisions with each other while crossing
the center.

Moreover, the range of initial orbital parameters compatible with our
assumed starting conditions obviously depends on the actual age of simulated clusters,
that cannot be univoquely defined (clusters of different ages and different
initial conditions may have the same apocenter and velocity we chose to start
simulations). However, we checked that in the adopted triaxial potential 
all GCs with masses in the range $(1.5$--$2.0) \times 10^7$ M$_\odot$ (see Section \ref{clumod} and Table
\ref{tab1}) and moving on either box orbits or loop orbits with pericenter $\simeq 0$
and initial apocenter in the interval $3.5$--$4$ kpc decay (due to dynamical friction)
to our starting conditions in about 1 Gyr (having set $M_b=3\times 10^9$ M$_\odot$ and
$r_b=200$ pc). The  decay time can be easily scaled to any GC mass, due to its inverse
linear proportionality to it, that means a decay time 2 Gyr for a GC half that
massive.

As regards the computational techniques, we adopted a direct $N$-body
representations for the stars in the clusters, simulating their
dynamics by mean of the parallel MPI `ATD' code \citep{bib2} whose main features
are resumed in Appendix~\ref{Numericalmethod}.


\subsection{The galactic model}
\label{bulgemodel}
We considered the same self-consistent triaxial model described in \citet{zeeuw} and
in PCV;
the same model was also used in CDM to study tidal
tails formation around clusters in absence of df and on not very elongated orbits.
It corresponds to a non-rotating ellipsoidal, triaxial distribution of matter with axial
ratios 2:1.25:1 (Schwarzschild, 1979), which leads to a projected profile in agreement with
that observed in spirals spheroids \citep[see, e.g.,][]{bertola91,matthews04} and in
elliptical galaxies \citep[see, e.g.,][]{wagner88,davies01,statler04}.
The potential produced can be expressed as the sum of a spherically symmetric
term due to a density following the modified Hubble's law
\beq
\rho_b(r)=\rho_{b0} \left[1+\left(\frac{r}{r_b}\right)^2\right]^{-3/2}, \label{hubble}
\eeq
with $r_b$ the core radius and $\rho_{b0}=M_b/r_b^3$, plus other two non-spherical
terms that give the triaxial behaviour, i.e.
\beq
\Phi(x,y,z)=A[\Phi_r(r)+\Phi_1(z,r)+\Phi_2(x,y,r)], \label{pot_bul}
\eeq
with $A\equiv 4\pi G M_b$ and
\begin{eqnarray}
\Phi_r&=&-\frac{1}{r}\ln\left(\frac{r}{r_b}+\sqrt{1+\left(\frac{r}{r_b}
\right)^2}\right), \label {phir}\\
\Phi_1&=&c_1\frac{3z^2-r^2}{2(r_b^2+c_2r^2)^{3/2}}, \\
\Phi_2&=&-3c_3\frac{x^2-y^2}{(r_b^2+c_4r^2)^{3/2}},
\end{eqnarray}
The coefficients $c_i$ have been chosen in such a way to have
density axial ratios roughly constant with $r$ \citep{zeeuw},
i.e.: $c_1=0.06408$, $c_2=0.65456$, $c_3=0.01533$, $c_4=0.48067$.
We refer to the parameter $M_b$ as the galaxy mass, though this
model yields a total infinite mass. Actually, $M_b\simeq
0.45M(r_b)$ being $M(r_b)$ the mass enclosed in the sphere with radius $r_b$.
The force produced by the galactic potential is evaluated analytically and
then added to each particle acceleration during the simulations.

With regard to the df, we used the generalization
of the Chandrasekhar formula \citep{chandra} to the triaxial case (PCV),
with a self-consistent evaluation of the velocity dispersion
tensor, taking also into account that the GC is an
extended object (see Appendix \ref{dynfric} for details).
We checked the dependence of df decay on applying it to the center
of mass of the GC and to its center of density (see Sect. 3.2).

\subsection{The cluster model}
\label{clumod}
Our $N$-body simulations involved four different clusters.
They sample a King distribution, with $M$ the total mass, $\sigma$
the central velocity dispersion, $r_t$ and $r_c$, the
limiting and the King radius respectively, $c=\log (r_t/r_c)$ the concentration
parameter and, finally, $t_c= (r_c^3/GM)^{1/2}$ the core-crossing time.
The `limiting radius' is the radius at which the King
distribution function drops to zero to model the presence of the
external field \citep{king66}.
Of course, for consistency, the limiting radius should be of the
same order of the tidal radius corresponding to the local tidal field.
Anyway, in order to model the presence of some tidal debris around the clusters
since the initial time, the limiting radius was chosen to be $20$--$60$\% larger
than the maximum tidal radius. This maximum value is achieved at the apocenter,
i.e. at the initial cluster position, where it can be estimated as
$r_0(M/M(r_0))^{1/3}\simeq 0.3$ (being $r_0\simeq 4$ and $M(r_0)\simeq 14$).
However, the amount of cluster mass lying outside the tidal radius is in any case
less than about 1\% of the total mass.
The initial values of the parameters are listed in Table \ref{tab1}.

Each cluster was represented with $N=5\times 10^5$ `particles' in the
numerical model.
The particles were assigned a mass according to a Salpeter's
mass distribution ($dN/dm\propto m^{-2.35}$) in the range
$3.3\times 10^{-11} \div 3.3\times 10^{-9}$
(if, e.g.,  $M_b=3\times 10^9$ M$_\odot$ then this range is
$0.1 \div 10$ M$_\odot$) then, all the particle masses were uniformly re-scaled 
such to give the desired $M$. 
The choice to represent the GC with the
\lq right\rq \ value for the total mass, thus overestimating the individual
star masses for a factor $N_{true}/N$, where $N_{true}$ is the real number of
stars in the GC, is the same done by, e.g., \citet{combes99}. Anyway, both
the relative stellar abundances and the total binding energy are correctly
reproduced, thus we expect that the representation of the external tidal
effects on the internal velocity distribution (which actually drives mass loss 
from the cluster) is correct as long as the spurious heating effect 
introduced by the overestimate of stellar masses is not relevant. 
As a matter of fact, we checked (see later
in this Sect. and Sect. 3.1.3) that this heating is absolutely negligible over the 
time length of our simulations. \\
There are, of course, other possibilites to pick particle masses. For example,
\citet{ideta} and \citet{dehn04} assume all particles to have the same mass. This corresponds
to an individual mass of about $130$ M$_\odot$ in the case of the
simulations of the formation
of the massive $\omega$ Cen cluster \citep{ideta}, and of about $1$ M$_\odot$ in the case of
the study of tidal tail formation around the light Pal 5 cluster. Another possible
choice is that of having particle masses distributed according to an assumed stellar
mass function, subsequently rescaling other relevant
GC characteristic parameters (as done by, e.g., \citet{baumg}). \\
All these choices have critical aspects, unavoidably due to
the limited number of particles allowed. In particular, as said above, 
our choice may imply spurious heating effects (as shown in the context 
of numerical galaxy formation by \citet{steinm}), in what fluctuations
over the mean field are enhanced by the too high \lq star\rq \ mass 
value.
However, fluctuations should not be exceedingly important because the total 
mass of the system (the most relevant parameter
in determining the GC mean field) is kept at its right value.
In this respect, note that the half-mass relaxation time $t_{rh}$ can be evaluated
as \citep{spitzer}:
\begin{equation}
\frac{t_{rh}}{t_b}=
\frac{t_c}{t_b}\times\frac{N}{7\ln \Lambda}, \label{trelax}
\end{equation}
where, e.g. for the most compact cluster \d, $t_c/t_b\sim 3.7\times 10^{-2}$
(Table \ref{tab1}).
Following \citet{gier}, $\Lambda \simeq 0.1N$, thus
the above written formula gives $t_{rh}\sim 240t_b$ for the system \d in the numerical
model. Since our simulations reached about $40t_b$, we can say that
`spurious' collisional effects should be always negligible.
We say `spurious' because, of course, such effects are even more negligible
for the `real', physical, clusters. Indeed, if, say, $M_b=3\times 10^9$ M$_\odot$,
then cluster \a, for instance, would have $M=2\times 10^7$ M$_\odot$; hence,
if the stellar average mass is $\mean{m}\simeq 0.3$ M$_\odot$, it would be
made up of $N=7\times 10^7$ stars, making the real relaxation time even
longer than the simulated time.
Finally, our simulations start with clusters that have an age less than their
internal 2-body relaxation time, so we assumed no initial mass segregation.

\begin{figure}
\plotone{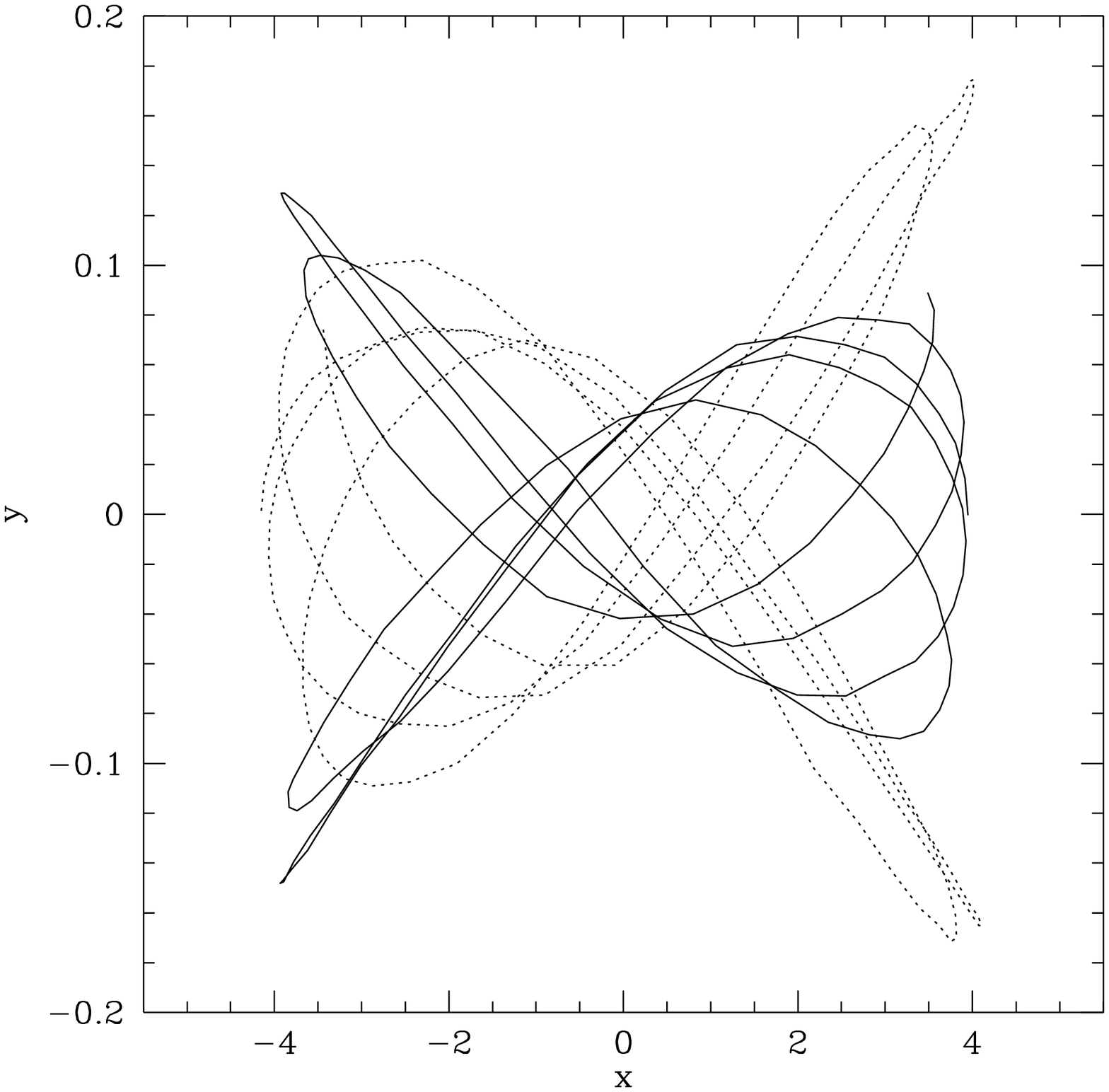}
\caption{Center-of-density trajectories on the $xy$ plane, for cluster \c
(dotted line) and \d (solid line) in the Simulation B. \label{orbit_c}}
\end{figure}

In Fig.~\ref{orbit_c} the orbits of the most concentrated clusters,
\c and \d are plotted, as turned out from simulation B (for which clusters
keep always a well defined core). They are quite
elongated box orbits and represent the motion of the cluster
\emph{\/center-of-density} (CD), i.e. the average of the
particles positions weighted with the local density instead of the mass
\citep{cashut}.
In most cases, we decided to take the CD
as the best suitable reference for that regards the study of the internal
cluster properties. Indeed, as we verified, the CM position is
strongly influenced by the very extended tidal tails which quickly formed,
so to be located even well outside the clusters' core.

\section{Results}
\label{results}

\subsection{The effects of the tidal shock on the GCs internal structure}

\subsubsection{The clusters internal `heating'}
\label{intheating}
\begin{figure}
\plotone{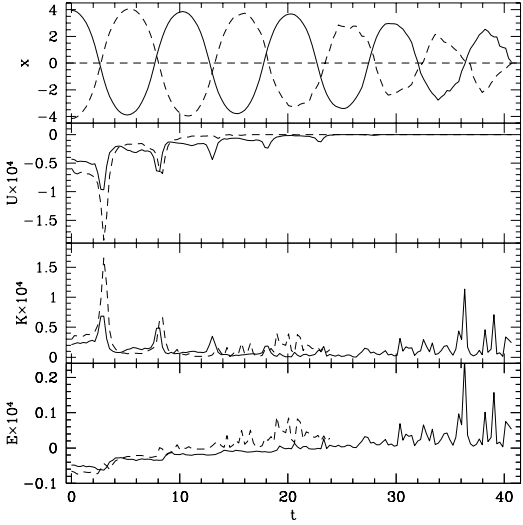}
\caption{From top to bottom: CD motion along $x$-axis for the pair
of (looser) clusters \a and \b in the simulation A;
time behaviour of the potential ($U$), kinetic ($K$) and
total ($E$) internal energy
(solid line: model \b, dashed: model \a).
Energies are evaluated in the center-of-density frame and they are concerned only
with the stars
in the sphere with radius equal to the initial King radius of each cluster.
For the loosest cluster \a the energy
is plotted up to $t\sim 24$ because, afterwards, it is almost destroyed.
The case of the cluster \a evolving alone in the
simulation C showed no appreciable differences in respect to
the simulation A.
\label{energiaA}}
 \end{figure}

\begin{figure}
\plotone{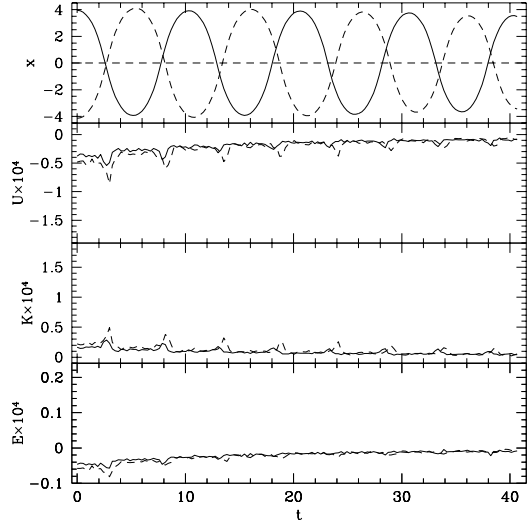}
\caption{From top to bottom: CD motion along $x$-axis for the pair
of the more compact clusters (\c and \d) in the simulation B, time behaviour of
the potential, kinetic and total internal energy
(solid line: model \d, dashed: model \c). Energies are evaluated as in
Fig.~\ref{energiaA}.
\label{energiaB}}
 \end{figure}

\begin{figure}
\plotone{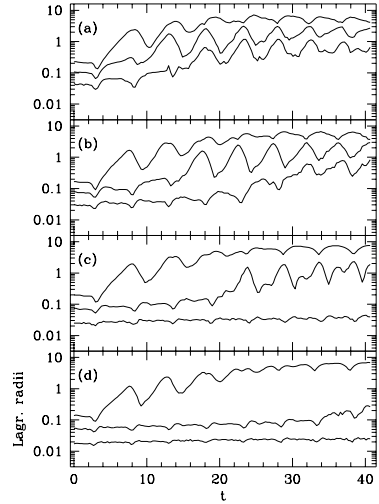}
\caption{Lagrangian radii evolution for the 4 cluster
models (as labelled). In each panel, the radii of
10, 50 and 90\% of the total mass are shown.
\label{rlag}}
 \end{figure}

In Fig.s \ref{energiaA} and \ref{energiaB}, for each cluster we plot the
total internal kinetic energy ($K$) and the internal gravitational potential
energy ($U$) of that part of the cluster that is enclosed in the sphere
centered at the CD with a radius equal to the initial King radius of the
cluster itself. `Internal' means that $K$ is evaluated with respect to the
cluster CM, while in $U$ the `external' potential produced by the galaxy and
by the other cluster is not taken into account.
Considering the least compact clusters, \a and \b,
one can note, from Fig.~\ref{energiaA},
that the potential energy drops off (whereas $K$ shows an `impulse') when the
clusters pass across the \emph{\/core} of the galaxy,
in agreement with the fact that their inner part undergoes a violent
compression at that moment, as clearly confirmed by
Fig.~\ref{rlag}.
Afterwards, the potential energy seems to recover its previous level,
but not completely because after each core-crossing a certain amount
of potential energy is irreversibly lost by the clusters, in favour of some
internal `heating'.
All this is true especially for the least compact clusters \a and \b, while
\c and \d models suffer this tidal influence in their external regions
only. Note, indeed, the much smaller variations in $U$ and $K$
(Fig.~\ref{energiaB})
and the nearly constant Lagrangian radius of 10\% of the mass
(Fig.~\ref{rlag}) for the most concentrated clusters.


The energy input is due to the tidal interaction with the environment
and not to the direct collision between the two clusters, as
verified by a careful comparison of the time behaviour of the $x$ coordinate of
the CDs and of the energy (Fig.s~\ref{energiaA} and \ref{energiaB}).
This is confirmed by the results of the simulation C that involves
the presence of the cluster \a \emph{\/alone}, thus excluding any effect due to
GC-GC collision. It can be affirmed that, at least within the
simulated time, both the energy behaviour and the motion of the cluster
show no appreciable difference when it is alone in comparison with the case
when it is in a pair.

Such a conclusion is not that surprising for orbits, like those
investigated here, that give rise to collisions with a high relative
velocity $V$ ($V/\sigma\sim 60$, being $\sigma\sim 0.1$ the cluster internal velocity
dispersion).
Notice, indeed, that the duration of a head-on collision is
$\sim r_c/V\sim 10^{-2}$ which is shorter than the crossing time of
our simulated clusters, so that the `impulse approximation' can be
applied in order to estimate the change in the clusters energy (per unit mass),
$E=U+K$, due to \emph{\/one} collision, that is \citep{bib4}:
$\Delta E\sim G^2M^2(V^2r_c^2)^{-1}\sim \sigma^4V^{-2}\sim |E|(\sigma/V)^2$. Hence
the number of collisions needed to change appreciably the internal energy of the
clusters is of order of $|E|/\Delta E\sim(V/\sigma)^2\sim 10^3$, i.e. much more
than that occurred in our simulated time. Of course, for
smaller $V$ such an approximation is no longer valid and the close
encounters may contribute furtherly to enhance the probability of merging
between clusters, as suggested by the simulations of superclusters merging
in \citet{fellhauer02} and discussed in detail in a forthcoming paper.

Note that while the loosest clusters \a and \b appear almost destroyed at $t\gtrsim 10$,
when $E\gtrsim 0$ and the CD oscillation around the galactic center begins to lose
its regularity (see Fig.~\ref{energiaA}), the more compact clusters in the simulation
B are able to survive the tidal stress (Fig.~\ref{energiaB}).

\subsubsection{The mass loss}
\label{themassloss}
The mass loss of the simulated GCs is due to the tidal interaction with
the galactic potential.
Indeed, we checked that the same cluster models used in the simulations if evolved at
rest and without any external field for a time equal to the time of the simulations,
exhibit a mass loss rate (in terms of particles escaping out of the limiting radius)
not greater than 1\%, presumably due to the (small) relaxation effects.

\begin{figure}%
\plotone{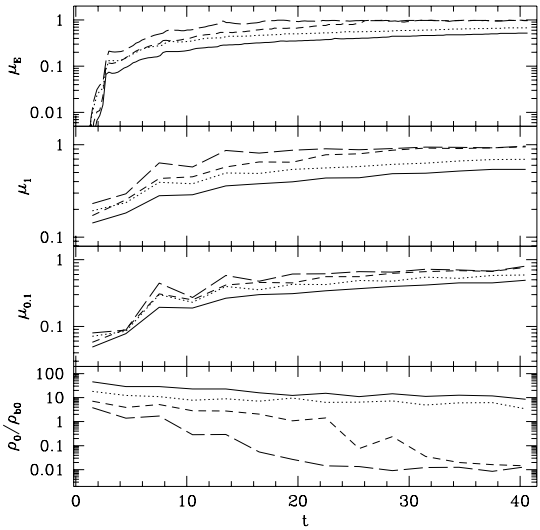}
\caption{Different evaluations of the time evolution of the fraction of mass lost
from the clusters (see text).
Bottom panel: time evolution of the central density normalized to the
galaxy central density. Solid line: model \d, dotted:
\c, short dashed: \b, long dashed: cluster \a in the simulation A.
The curves are time-averaged over $3t_b$.
\label{dens}}
 \end{figure}

We quantify the amount of the mass lost adopting various criteria. According
to an `energetic' one, $\mu_E$ is the fraction of the mass lost made up of
the stars whose internal energy ---i.e. that due only to the other stars of the same
GC and evaluated in the reference frame with the origin at the cluster CD--- is
non-negative. This criterion is not rigorous
because the sign of the individual star energy does not guarantee its
`dynamical destiny'.\\
Another indication of the mass lost from the clusters may be given with reference to
observations.
Defining as `lost' those stars that belong to regions where the local GC density
falls below $\alpha \rho_b$, the choices of $\alpha=1$ and $\alpha=0.1$ lead to the quantities
$\mu_1$ and $\mu_{0.1}$ as fractions (to the total initial ``observable'' GC mass)
of mass lost from the GC reported in Fig.~\ref{dens}.


Note that all the escaping criteria give very similar behaviours, especially
for the most compact clusters.
This behaviour is well fitted by a law of the form $\mu\sim 1-a\exp (-t/\tau)$;
in Table~\ref{massloss} the best-fit parameters are given for the various GCs
according to the different criteria adopted. Note, also, that although the initial
limiting radii are greater than the local tidal radius, the cluster
mass initially outside the tidal radius is, in any model, less
than the 1\% of the total mass, giving a negligible contribution
to the overall mass loss.

Even if the statistics is poor (only 4 models), nevertheless we
quantified a correlation between the King concentration parameter
$c$ and $\tau_E$ (the Pearson's linear correlation
coefficient between $\ln\tau_E$ and $\ln c$ is $0.992$).
The best power law correlation in the bi-logarithmic plane 
(least $\chi^2$ fit, corresponding to $\chi^2 =0.091$) is
$\tau_E/t_b\simeq 14\times c^{6.1}$.

\subsubsection{Density profiles and mass segregation}
\label{mass_segr}
\begin{figure}
\plotone{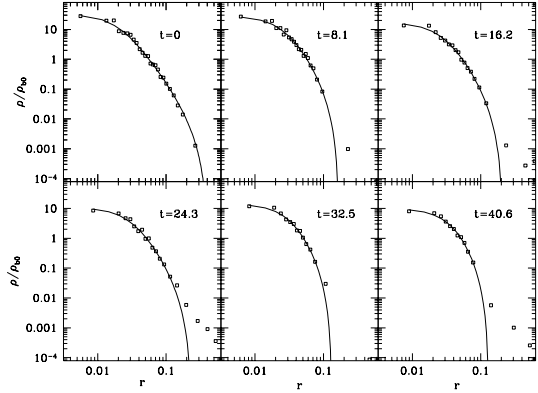}
\caption{Cluster density profile (squares), for the model \d at
various times. The sampled spherical shells have width such to contain a constant
number of particles. The density is normalized to the galaxy central density.
The solid curve is the best King model fit.
\label{rho_c01}}
 \end{figure}

\begin{figure}
\plotone{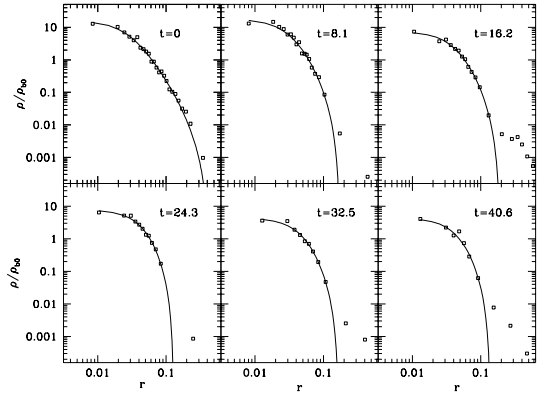}
\caption{As in Fig.~\ref{rho_c01} for the cluster \c.
\label{rho_c02}}
 \end{figure}

\begin{figure}
\plotone{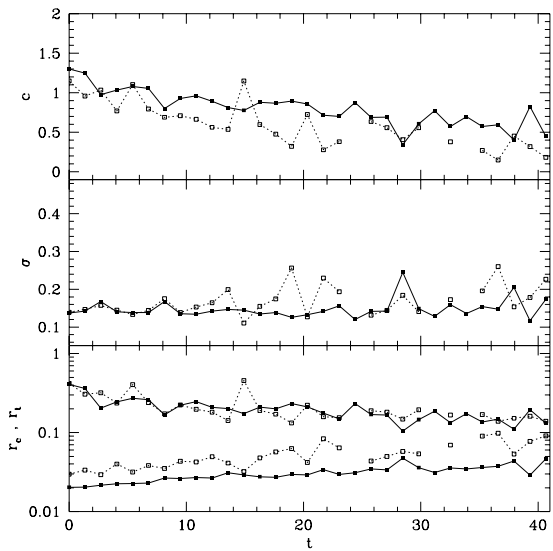}
\caption{From top to bottom: evolution of the concentration parameter, $c$, of the central
velocity dispersion, $\sigma$, and of the King and limiting radii ($r_c$, $r_t$) for
the best King model fits to clusters \c (open squares,
dotted line) and \d (solid squares and solid line).
\label{c_di_t}}
 \end{figure}

For the two most compact clusters (\c and \d) in the simulation B, we were
able to fit the inner radial density profile with a King distribution
throughout the whole simulated time, as shown in Fig.s \ref{rho_c01}--\ref{rho_c02}.
We also studied the time behaviour of the concentration parameter
($c$), of the central velocity dispersion ($\sigma$) and of the limiting ($r_t$)
and King ($r_c$) radii for both clusters (Fig.~\ref{c_di_t}).
Four points are of particular interest:
i) the GCs remain bound and keep a well defined spherical ``core'';
ii) the inner part of the clusters
tends towards less and less
\emph{\/concentrations} (see also the bottom panel of Fig.~\ref{dens});
iii) $c$ shows wide oscillations coinciding, temporally, with the repeated
compressions and re-expansions experienced by the clusters at each passage in
the galactic core (see Fig.~\ref{rlag}); iv) $\sigma$ appears to be
roughly constant.
The decrease of the concentration is due to both the growth of the
King radius and to the decrease of the limiting radius. The former is caused by
the continuous re-virialization of the core as the cluster loses mass, while the
latter is a direct consequence of the tidal erosion occurring preferentially
in the system outskirts.

It is worth noting that in our case the tidal erosion is strong enough
to act much more rapidly than 2-body collisions ($\tau \lesssim t_{rh}/10$).
This explains the apparent contradiction of our results, in particular point (ii),
with those of other authors investigating the role of tidal interaction, e.g.
\citet{gnedin99}.  Indeed, they found, by means of Fokker-Planck simulations
of a GC undergoing tidal forces much weaker than in our case,
that the tidal erosion acts just to accelerate the core collapse on its characteristic
time scale, which is of the order of tens of $t_{rh}$. At this same regard, we note that 
an initial growth of the core of not too concentrated clusters was
already found by \citet{spitzer73} and \citet{spitzer75} (in the cases of
single-mass and multimass Montecarlo models, respectively) on a time scale short
compared to $t_{rh}$, followed by a rapid evaporation-induced core collapse.

In order to investigate possible mass segregation
phenomena ---not expected to be relevant because the
simulated time is short in comparison with the 2-body
relaxation time---
we analysed the evolution of the average mass of stars within three
different clusters regions in simulation B:
only for cluster \c in the pair, a rather small increase of the average mass
is found ($\sim 10$\% of the initial value) in the inner part, i.e. in the sphere
where the cluster included initially 20\% of its mass.
In the outskirts and in the tails
no appreciable mass segregation occurs.
This confirms the absence of significant collisional effects that, if present,
would have indicated either an influence of the external tidal field in
decreasing the relaxation time, or a too small number
of particles with respect to the actual number of stars.
\subsection{The orbital decay}
\label{theorbitaldecay}
\begin{figure}
\plotone{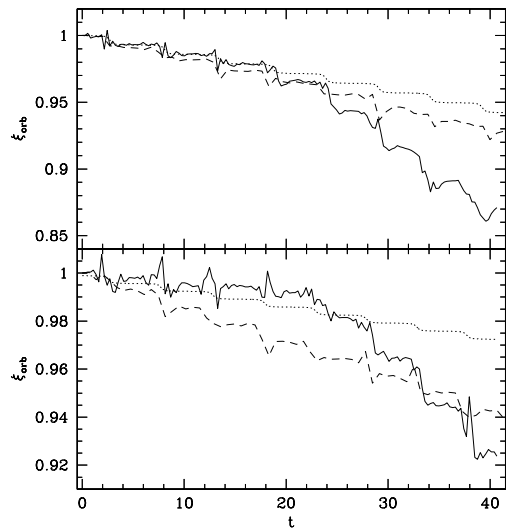}
\caption{Time behaviour of the fraction of orbital energy kept, at time t, by the
clusters in the simulation B (solid line) and by the corresponding point-mass
(dotted line). Upper panel: cluster \c; lower panel: cluster \d. For comparison, the
case with the df evaluated at the cluster CD by mean of the `reduced' simulations (see text)
is also plotted (dashed line).
\label{energiaBe}}
\end{figure}

In the framework of our main scientific motivation, it is important to quantify
how fast GCs decay towards the center of their parent galaxy.
At this purpose we defined
the adimensional quantity
\begin{equation}
\xi_{orb}(t)=\frac{E_{orb}(t)-\Psi_0}{E_0-\Psi_0},
\end{equation}
where $E_{orb}$ is the orbital energy of the CD of the cluster, $E_0=E_{orb}(0)$
and $\Psi_0=-4 \pi GM_b/r_b$ is the galactic potential well.
If neither dissipation nor
`energy injection' occur, then $\xi_{orb}=1$.
In Fig.~\ref{energiaBe} we give the time behaviour of $\xi_{orb}$ for the
clusters \c and \d, only, because they keep a rather defined core on the
whole duration of the simulation.
In order to understand whether the tidal interaction with the
galactic field influences appreciably the orbital decaying, it is
worth comparing the clusters $\xi_{orb}(t)$ evolution with that
of two point-masses, moving within the same
galactic potential
with the same initial conditions of the clusters and experiencing the same df
(Fig.~\ref{energiaBe}), but free from tidal effects.
Note that in the evaluation of the orbital energy of a cluster, the potential
energy given by the interaction with the other is not taken into account;
as we said in Sect.~\ref{intheating}, the overall evolution of a cluster is not
influenced appreciably by the presence of the other GC.
Thus, to simplify the comparison, we
considered the two point-masses as non-mutually interacting objects.
Otherwise, we would have had to smooth the mutual gravitational force with a suitably
variable smoothing radius, in order to simulate the variable tidal distortion
occurring for the clusters. Moreover, the mass of the point-masses is kept
constant for consistency with the choice of constant clusters half-mass as
a parameter in the df formula (see Appendix \ref{dynfric}).

We expected a different behaviour between the two cases, mainly because of
the tidal interaction with the galaxy, that occurs only in the full $N$-body case and that,
presumably, gives rise to a partial ``thermalisation'' of the orbital energy
among the internal degrees of freedom, resulting in a further form of dissipation.
This prediction is confirmed by the energy evolution, in the sense that the
\emph{\/rate} of dissipation is greater in the case of extended objects
than for the two point-masses.
The peaks shown by $\xi_{orb}$ are due to the
strong acceleration of the clusters CM during their close encounters
(the mutual potential energy is not taken into account).
To quantify the further frictional effect on the extended bodies, we followed the
decay of the point-masses up to the time $t_{dec}$ when $\xi_{orb}\simeq 10^{-2}$,
i.e. when their orbit is confined to a region of size comparable with that of a
typical GC. This happens at $t_{dec}\sim 400$ for both the point-masses. On the
other side, from Fig.~\ref{energiaBe}, we can extrapolate a value of $t_{dec}$
for the $N$-body systems assuming a constant energy decay rate, as $t_{dec}\lesssim 180$
for the cluster \c and $t_{dec}\lesssim 280$ for the more compact but less massive
cluster \d. Actually, these are overestimates, because the energy decay rate
is generally increasing with time \citep[see PCV and][]{cv04}.

We must note, however, that the energy dissipation of the clusters depends
on the particular way the df is evaluated. Indeed, as described in Appendix \ref{dynfric},
we used, in the Chandrasekhar formula, the kinematical quantities of the
clusters CM that, due to the quick formation of large tails of stripped material,
exibits a very rapid decay towards the center, where the density
and velocity dispersion of the galactic model have larger values than at the
location of the main body (core) of the cluster.
Actually, without an $N$-body self-consistent representation of the galaxy
in which the satellite moves, the way to compute and assign the df
deceleration to the bodies of a very extended object is not
a trivial issue. It is difficult to model correctly the df
changes induced by the cluster distorsion and, maybe, even more
difficult to take into account the gravitational feedback
of the cluster \emph{\/on} the galactic nuclear region, thus we were forced to do
unavoidable simplifications.
As explained in Appendix \ref{dynfric}, we assumed that the `global' effect
of the df is uniformly distributed to every GC star, and we evaluated it
as if the cluster were concentrated in the CM position with
the CM velocity.
Even if questionable, this is a logical choice that would deserve a discussion,
which is out of the purposes of this paper.
In any case, we re-simulated (though with $N=10^4$, for obvious 
computational convenience) the GCs evolution for models \c and \d (as individual systems 
in the galactic field), 
and with df computed by mean of the formula given by Eq. (\ref{a_df}) ``centered'' in
the CD instead of the CM, and the comparison is shown in Fig. \ref{energiaBe}.
The $\xi_{orb}$ behaviour in these new simulations is flatter than that of
the case with df evaluated at CM. The smaller effect of df is explained by 
that the CM is systematically closer to the galaxy center than the CD.
Nevertheless the different amount of df does not modify qualitatively the
global GC evolution at least within the simulated
time. Note that a reliable estimate of the mass loss in the df-on-CD case
cannot be achieved because the reduced number of particles ($N=10^4$) makes the
2-body relaxation time even shorter than the simulated time
(see Eq.~\ref{trelax}),
thus affecting significantly the evolution with (spurious)
collisional effects.

Considering the limited computational resources available, we chose to
adopt an as more accurate as possible $N$-body representation of the
cluster to study in detail the tidal disruption process, even if this
forced us to employ an analytic single-component model for the galaxy,
and so an analytic treatment of the df effect.
However, the presence of velocity anisotropy for the field stars
has to be take into account, since it is important in
altering the efficiency of the df as was proved by \citet{Bin77}
in the axisymmetric case, then by PCV in the triaxial case and recently
confirmed by the numerical simulations of \citet{penarrubia04}.
Thus, we decided to use the PCV generalisation of the Chandrasekhar formula, with
the self-consistent implementation
of the analytic expression for the velocity dispersion tensor provided
by the galactic model we adopted.
Moreover, we took into account the change in the df caused by
the non-uniform density of the field stars, through a suitable local
estimate of the Coulomb log (see Appendix \ref{dynfric}).
%
%
%

\subsection{Tidal tails morphology}
\begin{figure}
\plotone{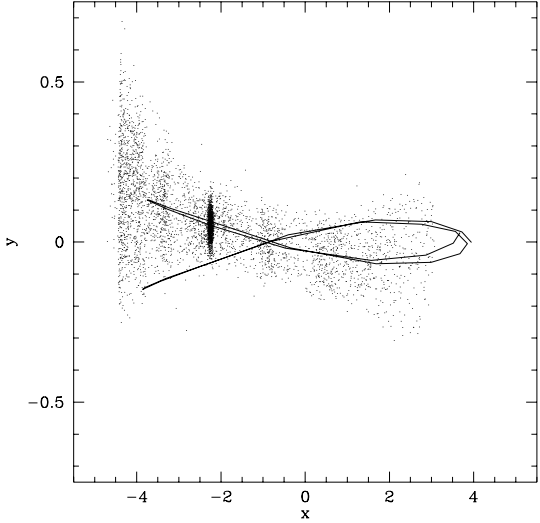}
\caption{Configuration of cluster \b at $t=17$. The projected trajectory of its
center-of-density motion is also plotted. The $y$-axis scale is expanded to
show better the alignment between the cluster orbit and the tails.
\label{tidalorbit}}
 \end{figure}

Tidal tails rapidly form around the clusters
and follow closely their orbit (see Fig.~\ref{tidalorbit}).
This tail-orbit alignement has been recently observed for the GC Palomar
5 \citep{oden01,oden03} and also reproduced in various simulations
\citep{laguna95,combes99,dehn04,nostro}.
Since the velocity dispersion in the cluster outskirts (where most of
the evaporated stars come from) is much lower than the
cluster orbital velocity, it is not surprising that,
there, stars move on orbits similar to that of the
cluster itself, as clearly shown by Fig.~\ref{tidalorbit}.
It can be also seen a significant spread of
escaped stars around the apocenter of the cluster orbit. This because of
the small differences among the velocities of the stars at the moment they leave
the cluster become unimportant at the apocenter, where the orbital velocity
is very low.
\subsubsection{Clumps and ripples formation}
In our case, where GCs orbits are quasi-radial, centrifugal and Coriolis'
forces are not important in determining
the tails shape distorsion investigated in CDM on a sample of less
elongated orbits.
Nevertheless, we do observe in our simulations the presence of stellar
overdensities along the tidal tails, similar to those seen around the mentioned
Pal 5 and in the simulations in CDM.

The cluster \b is plotted at various times close to the first passage at
the rightmost apocenter, in Fig.s \ref{l01_1}--\ref{l01_2}.
A ``ripple'' starts to form around $x=3.1$ at $t=8.6$, as a `wave-like'
overdensity.
Then, the cluster travels for a while across and then
above the formed ripple ($t=9.2$--$9.7$), before reaching its apocenter.
It can be also seen that at $t=10.3$ (Fig. \ref{l01_2}) another ripple
forms again around $x=4.1$, while at $t=9.7$ the one previously formed begins to
move inward giving rise, at $t=10.5$, to a clump
---an overdensity with a roughly spherical shape--- around $x=3.2$.
All these plots are projections on the $xy$ plane, however
all the structures are almost aligned along the orbit around which (i.e.
around the $x$-axis) they show a roughly cylindrical symmetry;
as an example the ripples are, actually, thin discoidal structures.
This can be seen in the available 3-d animations (see below).

To explain the overdensities found in the tails of the mentioned Palomar 5,
\citet{dehn04} raised an hypothesis stating that clumps could be due to
the effect of the interaction with Galactic substructures (like giant molecular clouds, spiral arms,
dark-matter sub-halos or massive compact halo objects). Basically, their opinion
is that small-scale overdensities like clumps can only be built up by the tidal
interaction with fluctuations on relatively small-scale of the external
gravitational field.
On the contrary, the results shown here seem to confirm the findings of CDM, in the
sense that in the absence of any small-scale substructure in the overall
potential clumps form too, thus suggesting that a further mechanism is at work.

Analysing the local velocity measured along the orbital path and the
energy of the stars in the cluster field, we can affirm that both clumps and the
apocenter ripple are not gravitationally-bound aggregates but, rather, overdensities
due to the local deceleration of the stellar motion, as already verified in CDM and
in \citet{celmech}. On the
contrary, where the velocity increases towards the direction of the stellar
mean motion along the tails, the tails tend to rarefy and an underdensity occurs.
The metaphor of the ``motorway traffic jam'' mechanism
used in  \citet{dehn04} is effective here, as can be seen in
Fig.~\ref{vmed}.
The cause of the overdensity is the strong 
deceleration of the stellar flux immediately on the left of the ripple location (around $x=3.25$).
Note that inside the cluster the profile of $\mean{v_x}$ is nearly flat as expected for
a bound (almost rigid) object, while in the ripple region it exhibits a
non-zero gradient.


It is interesting to compare a configuration like those shown in
Fig.~\ref{l01_2} at $t=10.3$ and $t=10.5$, with the `arcs' of material observed around
the spectacular shell galaxy NGC 3923 \citep{fort,pence}
as well as with the results of the simulations in  \citet{HQ1,HQ2,HS} and \citet{bourn}.
Although different scenarios and length-scale are involved, the qualitative
features of the debris produced by the tidal interaction of a compact and smaller
object with a larger density distribution, are rather similar to what we found
in this paper. In particular,
sharp-edged structures in form of `ripples' or `arcs' are formed when the
satellite moves on a very elongated orbit in a regular potential generated
by a much more massive and larger object.

To conclude this Section, we give the web address {\tt http://inaf.cineca.it/KP/kp4.html}
where the reader can get animations regarding some parts of the
simulation A.
The first animation refers to the whole simulation and one can see
how the least compact GCs get destroyed.
The second animation is a zoom around the apogalacticon, that corresponds to the
Figures \ref{l01_1}--\ref{l01_2} of the sub-structures formation.

\section{Conclusions}
\label{concl}

In this first of a series of papers we investigated the effects of the central galactic environment
on the structure of globular clusters decayed in the inner region because
of dynamical friction.
Our main concern is to answer the following questions: i) to what
extent are clusters able to survive the strong tidal interaction with the galaxy?
ii) do they keep their compact structure long enough to permit dynamical
friction to dissipate completely their orbital energy? iii) what are the effects of close
encounters and collisions between clusters?

The main conclusions of our work can be summarized as follows.

Sufficiently compact clusters (initial King concentration parameter $c\geq 1.2$) can
survive the tidal interaction with the external fields for, at least, the
duration of our simulations, i.e. for 8 passages across the galactic center
($t\simeq 40t_b$, being $t_b$ the galaxy core crossing time), while those with
$c=0.8$--$0.9$ are almost disgregated after only $2$--$3$ passages ($\sim
10$--$15 t_b$). The mass loss from the clusters as
a function of time follows an exponential law $\sim e^{-t/\tau}$ with $\tau$ up
to $\sim 70t_b$ for the most compact cluster ($c=1.3$). Even if still without a high
statistical reliability, we found a correlation between the mass loss time-scale
and the concentration parameter that
can be quantified as $\tau\simeq 14c^{6.1}$. This means that a GC with an initial
$c\gtrsim 1.6$ keeps bound a substantial amount of its
mass up to $t\simeq 300t_b$, i.e. up to the complete orbital decaying.

During their evolution, the survived clusters maintain the initial King profile
in the inner region, with a \emph{\/decreasing} concentration and
a \emph{\/constant} central velocity dispersion.
A small degree of mass segregation occurs at the end of
the simulations: the average stellar mass in the central clusters
region increases of about 10\% of the initial value, but the fluctuations are
even larger. The very low level of mass segregation is not surprising,
for the half-mass relaxation time is $\sim 6$ times longer than the simulated time.

With regard to the orbital decay, we found that the tidal interaction
with the field gives rise to a dissipation of the orbital energy with a rate
comparable to that given by the df,
when this is evaluated at the cluster CM. However,
even when the CD is used instead of this latter, we found that the tidal
interaction provides a further important mechanism of orbital decaying
besides dynamical friction (included in our simulations).
This is important for the validity
of the nucleus accretion model according to which the nuclear region receives
a large amount of mass in form of clusters that have lost their orbital
energy \citep{bib3,cv04}.

Another relevant finding of this work is that, in the cases considered here, the tidal
interaction between the two clusters, even during face-on collisions, produces
negligible effects on both their
internal evolution and their orbital evolution (at least for $t<40t_b$), 
confirming the estimates done according to the impulse approximation.

All these results indicate that sufficiently massive and compact clusters
can survive the disruptive effect due to the strong tidal forces exerted
by the central environment. Dynamical friction \emph{\/and} tidal dissipation
are then able to brake many globular clusters \emph{\/before} they are
disgregated, so to allow the formation of a dense and massive
super-cluster resulting from merging events among clusters.
The simulations shown here did not achieve the final merging stage
between the objects, because of the limits in the computational resources.
In a forthcoming paper, we will present results of
a new series of simulations directly regarding the merging among clusters.
Indeed, the importance of the present work is also to get physically reliable
and realistic initial conditions from which the simulations of the
final merging process will start.

A further result of this work concerns the formation and structure
of tidal tails.
Tidal tails form rather quickly, with a clear tendency to align along
the orbital path. They exhibit complex morphology, with the
presence of clumps and `ripples' as those observed around globular clusters
and shell galaxies, too. Such overdensities are not gravitationally bound
aggregates but, rather, their formation seems to have a kinematical origin, being
connected to the deceleration of the stellar `flux' motion along the tails.


\section{Acknowledgements}
The main computational resources employed for this work
were provided by CINECA (http://www.cineca.it) and INAF (http://inaf.cineca.it)
agreement under the \emph{\/Key-Project} grant \emph{\/inarm033}.
Part of this work was also supported by MIUR (Ministero dell'Istruzione
dell'Universit\'a e della Ricerca)
under the \emph{\/PRIN 2001} funding program
(grant 2001028897\_005).

\appendix

\section{Treatment of the dynamical friction}\label{dynfric}
The effect of braking due to df on globular
clusters was followed by the same approach used in PCV.
The classic formula developed by \citet{chandra} for the df deceleration term has been
extended by PCV to the triaxial case, in partial analogy with the \citep{Bin77}
extension to the axisymmetric case, obtaining:
\begin{equation}
\mathbf{a}_{df}=-\gamma _{1} V_{1} \mathbf{\widehat{e}}_1-\gamma _{2} V_{2}
\mathbf{\widehat{e}}_2-\gamma _{3} V_{3} \mathbf{\widehat{e}}_3
\label{a_df}
\end{equation}

where $\mathbf{\widehat{e}}_i$ ($i=1,2,3$) are
the eigenvectors of the velocity dispersion tensor of the galaxy stars
and $V_i$ is the component of the velocity of the baricenter of the GC along
the $\mathbf{\widehat{e}}_i$ axis.
The coefficients $\gamma_i$ are (see PCV):
\begin{eqnarray}
\label{gammai}
\gamma_i&=&\frac{2\sqrt{2\pi}\rho (\mathbf{r})G^{2}M\ln \Lambda}
{\sigma _{1}^{3}} \times\\
\nonumber & & \times \int_{0}^{\infty} {\exp\left(-\sum_{k=1}^{3}
\frac{V_k/2\sigma_{k}^{2}}{\epsilon _{k}^{2}+u}\right)}
\times  \\
\nonumber && \times{(\epsilon _{i}^{2}+u)^{-1}\left[\sum_{k=1}^3 (\epsilon _{k}^{2}+u)\right]^{-1/2}}du
\end{eqnarray}

where: $\rho(\bf r)$ is the mass density of background stars,
$\ln \Lambda$ is the Coulomb's logarithm, $M$ is the mass of the test object, $G$
is the gravitational constant,
$\sigma_i$ is the eigenvalue, corresponding to
$\mathbf{\widehat{e}}_i$, of the velocity dispersion tensor $\sigma_{ij}$
evaluated in $\mathbf{r}$,
and $\epsilon_i$ is the ratio between $\sigma_i$ and the greatest eigenvalue, set as $\sigma_1$.

The velocity dispersion
was computed and presented by \citet{M80} for the Schwarzschild ellipsoid,
both in the case of a rotating and a non-rotating model. In this paper
we considered just the non-rotating model.

For computational convenience, the Merritt's data were fitted obtaining three analytical expressions for the eigenvalues
$\sigma_3 < \sigma_2 < \sigma_1$ at $\tilde r\equiv |\mathbf{r}|/r_b$:
\begin{eqnarray}
\label{autoval}
\nonumber && \sigma_1^2=3.1 e^{-\tilde r/9.2}\\
&& \sigma_2^2=\frac {2.9} {1+0.43 \tilde r^{1.60}} \\
\nonumber && \sigma_3^2=\frac{2} {1+0.44 \tilde r^{1.7}}
\end{eqnarray}
(expressed in unit of $GM_b/r_b$). Analogously, the following fitting formulas for the Euler angles giving the orientation
of the local reference frame where $\sigma_{ij}$ is diagonal were determined:
\begin{eqnarray}
\label{autovet}
\nonumber &&\alpha = \pi/2+\frac{0.2\tilde r^{2.3}} {1.+0.2 \tilde r^{2.3}}\left[\arccos (\tilde z/\tilde r)-\pi/2\right]  \\
&& \beta= \pi/2 + \frac {0.2\tilde r^{2.25}} {1+0.2\tilde r^{2.25}}\arcsin\left(\frac {\tilde y}{\sqrt{\tilde x^2+\tilde y^2}}\right) \\
\nonumber && \gamma= \frac {0.8\tilde r^3} {1+0.8\tilde r^3}\arcsin\left(\frac {\tilde y}{\sqrt{\tilde x^2+\tilde y^2}}\right)
\end{eqnarray}
The fitting formulas \ref{autoval} and \ref{autovet} are a slight improvement of those reported in PCV.

Given $b_M$ and $b_m$ respectively,
the maximum and the minimum impact parameter of the test object with
field stars, we adopted a variable and local Coulomb's logarithm $\ln \Lambda=\ln (b_M/b_m)$ assuming
$b_m= GM_{1/2}/\sigma^2(\mathbf{r})$, where $\sigma^2(\mathbf{r})=\sigma_1^2 + \sigma_2^2 + \sigma_3^2$
and $M_{1/2}=M/2$ is the cluster initial half-mass.
Indeed, since the GC is not a point-mass,
the df is made less efficient by the `weakening' of the encounters with
impact parameters smaller than the size of the object itself, assumed as
the half-mass radius \citep[see e.g.][]{bib4}.
Of course, this reasoning is based on the hypothesis that at least the core of the cluster
survives to the tidal stress so to make df to be able to continue acting on a compact object
with mass $\sim M_{1/2}$, throughout the whole simulated time.
Our choice of letting $b_M=10r_b$ seems quite adequate to our quasi-radial orbits,
noting that the stellar density at a distance
$r>10r_b$ from the galactic center falls below $10^{-3}$ times the central value.
Choices of $b_M$ as function of the galactocentric distance, like for instance
$b_M=r$ \citep{hashimoto03}, are of not straightforward use in our cases.

Finally, the acceleration given by Eq. (\ref{a_df}) is
evaluated either at the CM or at the CD position of the cluster and then added to the
accelerations of every particle of the cluster.

\section{The $N$-body approach}\label{Numericalmethod}
The simulations have been performed by means of the code `ATD' \citep{bib2}. It is a
a parallel \emph{\/tree-code} that follows, in part, the \citet{bh} algorithm.
For the potential due to distant `particles' it uses a multipolar expansion truncated at the
quadrupole moment and has been parallelized to run on high performance computers
via MPI routines, employing an original parallelization approach.
A good resolution in the evaluation of the total gravitational force acting on
each star of the cluster is achieved by a direct summation of the contribution
given by neighbours. To avoid instability in the time-integration, the $1/r$ gravitational
potential has been smoothed by using a continuous $\beta$-spline function
for $r < \epsilon$, such to give an exactly Newtonian potential
for $r > \epsilon$ \citep{bib5} and a force that vanishes as $r$ for $r \to 0$.
The time-integration of the `particles' trajectories is performed according to the
leap-frog algorithm. This latter uses individual and
variable time-steps according to the block-time scheme \citep{aars,bib5}, in
addition with corrections we implemented in order to keep the same order of
approximation also during the time-step change. Denoting by $t_{\rm{c}}$ the
minimum core-crossing time of our simulated clusters, the maximum allowed
time-step is $\Delta t_{\rm{max}} =0.01 t_{\rm{c}}$, while the minimum is
$\Delta t_{\rm{min}} = \Delta t_{\rm{max}} / 2^{10}$, thus fastest particles may
have a time-step as small as $\sim 10^{-5} t_{\rm{c}}$. To choose the time-step
of the $i$-th particle we found that the best
compromise between accuracy and speed in the time-integration is via
the formula: $\Delta t_i=0.05\times \min\{(d_i / a_i)^{1/2}, d_i / v_i\}$, where $v_i$ is the
velocity of the particle relative to its first neighbour (or to the CM of
its closest box), $d_i$ its distance from the first neighbour and $a_i$ the
modulus of the total acceleration of the $i$-th particle itself.

In all the runs the softening length $\epsilon$ is set to $10^{-5}r_b$ (i.e. $2\times 10^{-3}$ pc if
$r_b=200$ pc), so to
have $(\epsilon^3 / GM)^{1/2} \sim \Delta t_{\rm{min}}$.
This is a minimal condition to ensure that the level of accuracy in the forces
evaluation agrees with the accuracy in the time-integration scheme.
Note that such an $\epsilon$ is much smaller than the typical interparticle
distance so to keep the overall correct Newtonian behaviour.
For $N$ lower than the real number of stars, as in our case, the choice of the
gravitational smoothing radius is a matter of compromise between the
need for mantaining the Newtonian behaviour (the smaller is $\epsilon$
the more correct is the force evaluation) and the need for avoiding
a spurious increase of collisionality (the larger is $\epsilon$
the closer is the model relaxation time to the real one).As a matter of
fact, the results discussed in Sect.~\ref{mass_segr} suggest 
that $\epsilon$ is large enough not to introduce fictitious collisional 
effects.


\begin{figure}
\plotone{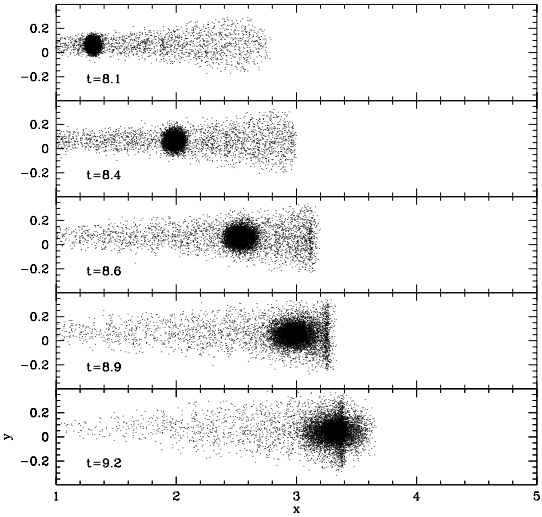}
\caption{Cluster \b is sketched at various times (as labelled) before the first
passage to the rightmost apocenter (galactic center is at $x=0$).
\label{l01_1}}
 \end{figure} 
\begin{figure}
\plotone{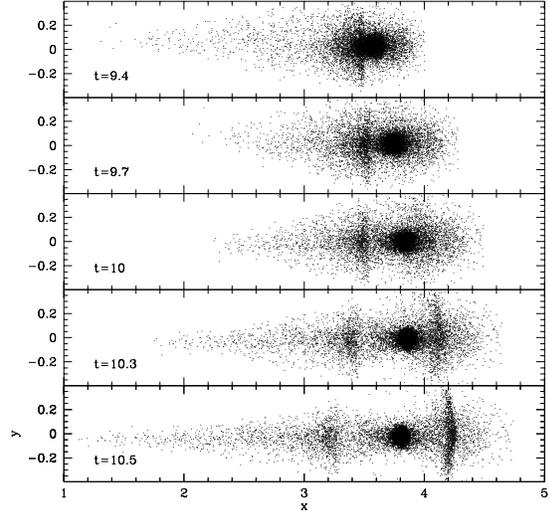}
\caption{Continuation of Fig.~\ref{l01_1} for the same cluster \b; the apocenter
is reached at $t\sim 10$.
\label{l01_2}}
 \end{figure} 
\begin{figure}
\plotone{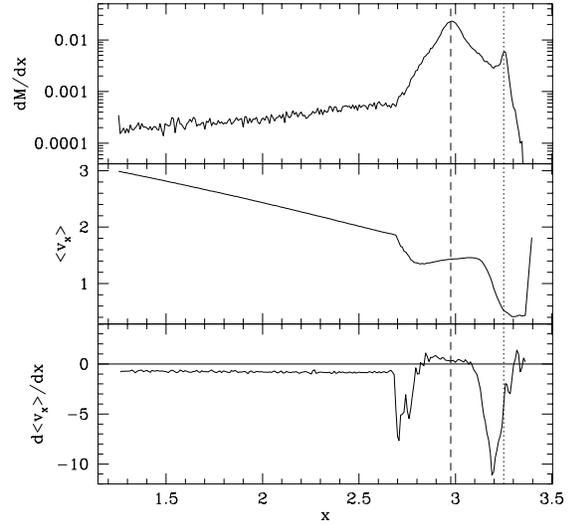}
\caption{The plots refer to the cluster shown in Fig.~\ref{l01_1} at $t=8.9$.
Upper panel:
linear density as measured along $x$-axis. Middle panel: averaged stellar
velocity component ($\mean{v_x}$) along $x$. Lower: spatial derivative of $\mean{v_x}$ (the
horizontal line indicates the zero value). The position of the GC and of
the `ripple' are marked, respectively, by the dashed and
dotted lines (galactic center is at $x=0$).
\label{vmed}}
 \end{figure} 
\begin{figure}
\plotone{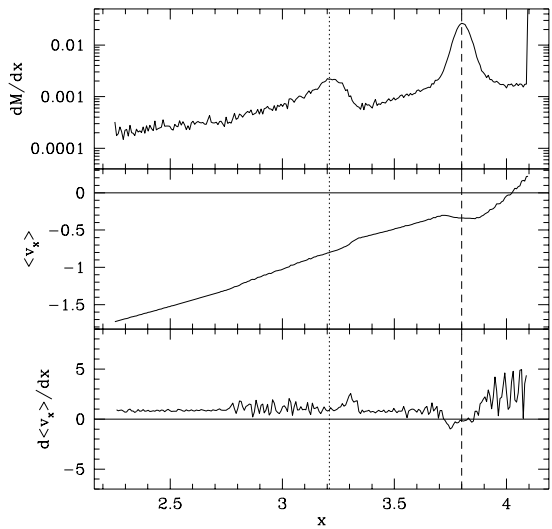}
\caption{As in Fig.~\ref{vmed} but referring to the time $t=10.5$.
The position of the GC is marked by the dashed line, while the `clump'
is at the dotted one.
\label{vmed2}}
 \end{figure} 

\begin{deluxetable}{p{0.8cm}ccccccp{1.3cm}}
\tablecaption{Clusters initial parameters.
\label{tab1}}
\tablewidth{0pt}
\tablehead{
\colhead{cluster model} & \colhead{$M\times 10^3$} &\colhead{$r_t$} & \colhead{$c$} & \colhead{$r_c\times 10^2$}
 & \colhead{$t_c\times 10^2$} & \colhead{$\sigma$} & \colhead{Simulation}
}
\startdata

{\bf a}&
$6.7$&
$0.47$&
$0.8$&
$7$&
$23$&
$0.13$&
{\bf A}, {\bf C}\\

{\bf b}&
$5$&
$0.36$&
$0.9$&
$4.5$&
$13$&
$0.13$&
{\bf A}\\

{\bf c}&
$6.7$&
$0.49$&
$1.2$&
$2.7$&
$5.4$&
$0.14$&
{\bf B}\\

{\bf d}&
$5$&
$0.38$&
$1.3$&
$1.9$&
$3.7$&
$0.14$&
{\bf B}\\
\enddata

\tablecomments{Parameters list for the initial cluster models.
Masses, lengths and time are in unit of $M_b$, $r_b$ and $t_b$
respectively.
}
\end{deluxetable}


\begin{deluxetable}{p{2.cm}cccccc}
\tablecaption{Parameters for the mass loss fits.
\label{massloss}}
\tablewidth{0pt}
\tablehead{
\colhead{cluster model} & \colhead{$a_E$} & \colhead{$\tau_E$} & \colhead{$a_{0.1}$} & \colhead{$\tau_{0.1}$}
 & \colhead{$a_{1}$} & \colhead{$\tau_{1}$}
}


\startdata

{\bf a} (sim. A)&
$3.8$&
$3.2$&
$0.75$&
$35$&
$0.51$&
$15$\\

{\bf b}&
$1.9$&
$9.7$&
$0.91$&
$33$&
$1.2$&
$14$\\

{\bf c}&
$0.77$&
$46$&
$0.83$&
$56$&
$0.72$&
$45$\\

{\bf d}&
$0.89$&
$65$&
$0.88$&
$74$&
$0.79$&
$69$\\
\enddata

\tablecomments{Parameters of the mass loss evolution $1-ae^{-t/\tau}$,
according to the criteria based on: the total internal energy ($a_E,\tau_E$),
the density contrast of $10$\% ($a_{0.1},\tau_{0.1}$) and $100$\% ($a_{1},\tau_{1}$).
Time is in unit of $t_b$.}

\end{deluxetable}

\end{document}